\documentclass[12pt,preprintnumbers,amsmath,amssymb,floatfix,prb,superscriptaddress]{revtex4}
\usepackage{graphicx}
\usepackage{epstopdf}
\usepackage{dcolumn}
\usepackage{bm}
\usepackage{longtable}
\usepackage{graphics}
\usepackage{amssymb}
\usepackage{amsmath}
\usepackage{xspace}
\usepackage{epsfig}
\newcommand {\etal}{{\it et al}. }

\newcommand{\eq}[1]{Eq. (\ref{eq:#1})}

\begin{document}
\title{Unified explanation of the Kadowaki-Woods ratio in strongly correlated metals}
\author{A. C. Jacko}
 \affiliation{Centre for Organic Photonics and Electronics, School of Physical Sciences, University of Queensland, Brisbane, Queensland 4072, Australia}
\author{J. O. Fj\ae restad}
\affiliation{School of Physical Sciences, University of Queensland, Brisbane, Queensland 4072, Australia}
\author{B. J. Powell}
\email{bjpowell@gmail.com} \affiliation{Centre for Organic Photonics and Electronics, School of Physical Sciences, University of Queensland, Brisbane, Queensland 4072, Australia}


\maketitle

{\bf Discoveries of ratios whose values are constant within broad classes of materials have led to many deep physical insights. The Kadowaki-Woods ratio (KWR)\cite{Rice,KW} compares the temperature dependence of a metalÕs resistivity to that of its heat capacity; thereby probing the relationship between the electron-electron scattering rate and the renormalisation of the electron mass.
 However, the KWR takes very different values in different materials\cite{Hussey,Dressel}.   Here we introduce a ratio, closely related to the KWR, that includes the effects of carrier density and spatial dimensionality and takes the same (predicted) value in organic charge transfer salts, transition metal oxides, heavy fermions and transition metals - despite the numerator and denominator varying by ten orders of magnitude. Hence, in these materials, the same emergent physics is responsible for the mass enhancement and the quadratic temperature dependence of the resistivity and no exotic explanations of their KWRs are required.}

In a Fermi liquid the temperature dependence of the electronic contribution to the heat capacity is linear, i.e., $C_{el}(T)=\gamma T$. Another prediction of Fermi liquid theory\cite{pinesandnoz} is that, at low temperatures, the resistivity varies as
$\rho(T)=\rho_0 + A T^2 $.
This is observed experimentally when electron-electron scattering, which gives rise to the quadratic term, dominates over electron-phonon scattering. 

\begin{figure}[p]
\epsfig{file=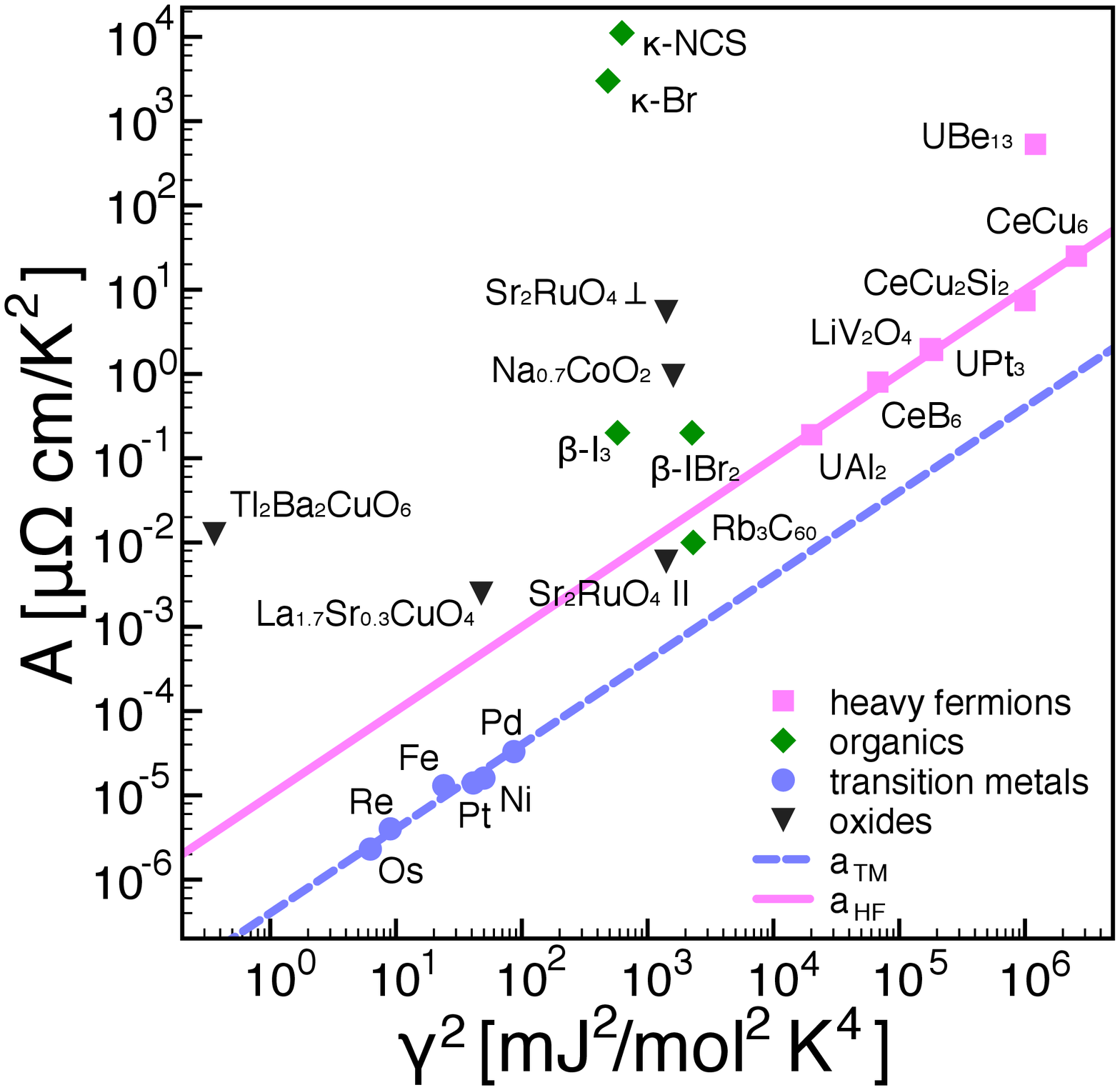,angle=0,width=8cm} 
\caption{
 The standard Kadowaki-Woods plot. It can be seen that the data for the transition metals and heavy fermions (other than UBe$_{13}$) fall onto two separate  lines. However, a wide range of other strongly correlated metals do not fall on either line or between the two lines. $a_{TM}$\,=\,0.4\,$\mu\Omega$\,cm\,mol$^2$\,K$^2$/J$^2$ is the  value of the KWR observed in the  transition metals\cite{Rice} and $a_{HF}$\,= 10\,$\mu\Omega$\,cm\,mol$^2$\,K$^2$/J$^2$ is the value seen in the heavy fermions.\cite{KW} In labelling the data points we use the following abbreviations: $\kappa$-Br is $\kappa$-(BEDT-TTF)$_2$Cu[N(CN)$_2$]Br; $\kappa$-NCS is $\kappa$-(BEDT-TTF)$_2$Cu(NCS)$_2$; $\beta$-I$_3$ is $\beta$-(BEDT-TTF)$_2$I$_3$; and $\beta$-IBr$_2$ is $\beta$-(BEDT-TTF)$_2$IBr$_2$. For Sr$_2$RuO$_4$ we show data for $A$ measured with the current both perpendicular and parallel to the basal plane, these data points are distinguished by the symbols $\perp$ and $\|$ respectively.  Further details of the data are reported in the supplementary information.}
\label{fig:KWR-trad} 
\end{figure}

 Rice observed\cite{Rice} that in the transition metals $A/\gamma^2\approx a_{TM}=0.4~\mu\Omega$\,cm\,mol$^2$\,K$^2$/J$^2$ (Fig.~\ref{fig:KWR-trad}), even though $\gamma^2$ varies by an order of magnitude across the materials he studied. Later, Kadowaki and Woods\cite{KW} found that in many heavy fermion compounds $A/\gamma^2\approx a_{HF}=10~\mu\Omega$\,cm\,mol$^2$\,K$^2$/J$^2$ (Fig.~\ref{fig:KWR-trad}), despite  the large mass renormalisation which causes $\gamma^2$ to vary by more than two orders of magnitude in these materials. Because of this remarkable behaviour $A/\gamma^2$ has become known as the Kadowaki-Woods ratio. However, it has long been known\cite{KW,MMV} that the heavy fermion material UBe$_{13}$ has an anomalously large KWR. More recently, studies of other strongly correlated metals, such as the transition metal oxides\cite{Li,Hussey} and the organic charge transfer salts\cite{Dressel,JPCMrev}, have found surprisingly large KWRs (Fig.~\ref{fig:KWR-trad}). It is therefore clear that the KWR is not the same in all metals; in fact it varies by more than seven orders of magnitude across the materials shown in Fig.~\ref{fig:KWR-trad}.

Several important questions need to be answered about the KWR: (i) Why is the ratio approximately constant  within the transition metals and within the heavy fermions (even though many-body effects cause large variations in their effective masses)? 
(ii) Why is the   KWR  larger for the heavy fermions than it is for the transition metals? And (iii) why are such large and varied KWRs observed in layered metals such as the organic charge transfer salts and transition metal oxides? The main aim of this paper is to resolve question (iii). We will also make some comments on the first two questions, which have been extensively studied previously.

\begin{figure}[p]
\epsfig{file=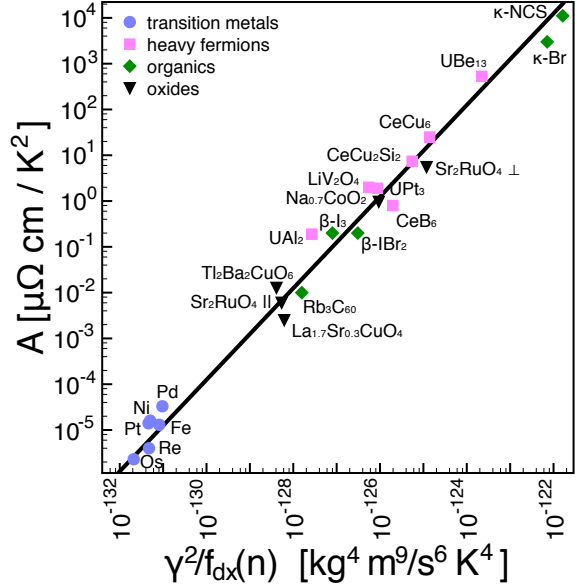,angle=0,width=8cm}
\caption{
 Comparison of the ratio defined in Eq. (\ref{eqn:newRatioGen}) with experimental  data. It can be seen that, in all of the materials studied, the data are in excellent agreement with our prediction (line).   The abbreviations in the data point labels are the same as in Fig.~\ref{fig:KWR-trad}.  Further details of the data are given in the supplementary information.   }
\label{fig:KWR-new}
\end{figure}

There have been a number of studies of the KWR based on specific microscopic Hamiltonians (see, for example, Refs. \onlinecite{Yamada,Auerbach,Coleman,LiR,Kontani}). However,  if the KWR has something general to tell us about  strongly correlated metals, then one would also like to understand which features of the ratio transcend specific microscopic models. Nevertheless, two important points emerge from these microscopic treatments of the KWR: (a) if the  momentum dependence of the self-energy can be neglected then the many-body renormalisation effects on $A$ and $\gamma^2$ cancel and (b) material specific parameters are required in order to reproduce the experimentally observed values of the KWR. Below we investigate the KWR using a phenomenological Fermi liquid theory; this work builds on previous studies of related models\cite{MMV,Hussey}. Indeed our calculation is closely related to that in Ref. \onlinecite{MMV}. The main results reported here are the identification of a ratio [Eq.~(\ref{eqn:newRatioGen})] relating $A$ and $\gamma^2$, which we predict takes a single value in a broad class of strongly correlated metals,  and the demonstration that this ratio does indeed describe the data for a wide variety of strongly correlated metals (Fig.~\ref{fig:KWR-new}).

It has been argued that the KWR is larger in the heavy fermions than the transition metals because the former are more strongly correlated (in the sense that the self-energy is more strongly frequency dependent) than the latter\cite{MMV}. 
Several scenarios have been proposed to account for the large KWRs observed in UBe$_{13}$, transition metal oxides and organic charge transfer salts including impurity scattering\cite{MMV}, proximity to a quantum critical point\cite{Li} and the suggestion that electron-phonon scattering in reduced dimensions might give rise to a quadratic temperature dependence of the resistivity\cite{Strack}. Hussey\cite{Hussey} has previously observed that using volumetric (rather than molar) units for $\gamma$ reduces the variation in the KWRs of the transition metal oxides. However, even in these units, the organic charge transfer salts have KWRs orders of magnitude larger than those of other strongly correlated metals.   We will argue that the different KWRs observed across this wide range of materials  result from the simple fact that the KWR contains a number of material specific quantities. As a consequence, when one replaces the KWR with a ratio that accounts for these material specific effects [Eq.~(\ref{eqn:newRatioGen})] the data for all of these materials does indeed lie on a single line (Fig.~\ref{fig:KWR-new}).

Many properties of strongly correlated Fermi liquids can be understood in terms of a momentum independent self-energy\cite{DMFT,Hewson}. Therefore, following Ref. \onlinecite{MMV}, we assume that the imaginary part of the self-energy, $\Sigma''(\omega,T)$, is given by
\begin{eqnarray}
\Sigma''(\omega,T) = -\frac{\hbar}{2 \tau_0}- s 
\frac{ \omega^2 + \left(\pi k_B T\right)^2}{\omega^{*2}}\label{eq:selfenergy} 
\end{eqnarray}
for  $| \omega^2 + \left(\pi k_B T\right)^2 | <\omega^{*2} $ and 
$\Sigma''(\omega,T) = - [{\hbar}/{2 \tau_0}+s
] F( [ { \omega^2 + \left(\pi k_B T\right)^2}]^{1/2}/ {\omega^{*}} )$ 
for  $| \omega^2 + \left(\pi k_B T\right)^2 | > \omega^{*2}$,
where $2s/\hbar$ is the scattering rate due to electron-electron scattering in the absence of quantum many-body effects, $\tau_0^{-1}$ is the impurity scattering rate, $F$ is a monotonically decreasing function with boundary conditions $F(1) = 1$ and $F(\infty) = 0$, and $\omega^*$ is determined by the strength of the many-body correlations. (See the Methods section for further discussion of the self energy.)

The diagonal part of the conductivity tensor may be written as\cite{mahan}\begin{equation}
\sigma_{xx}(T) = \hbar e^2 \int \frac{d\bm{k}}{(2\pi)^3}v_{x0}^2 \int \frac{d \omega}{2\pi} \mathcal{A}^2(\bm{k},\omega)\left( \frac{-\partial f(\omega)}{\partial \omega}\right), \label{eq:sigmakubo}
\end{equation}
where $v_{x0}=\hbar^{-1}\partial\varepsilon_{0}(\bm{k})/\partial k_x$ is the unrenormalised velocity in the $x$ direction, $f(\omega)$ is the Fermi-Dirac distribution,
$\mathcal{A}(\bm{k},\omega) = -2\,\text{Im}\{[\omega - \varepsilon_0(\bm{k}) +\mu^* - \Sigma(\omega,T)]^{-1}\}$ is the spectral density,
$\varepsilon_0(\bm{k})$ is the non-interacting dispersion relation and $\mu^*$ is the chemical potential. 
Note that Eq. (\ref{eq:sigmakubo}) does not contain vertex corrections; the absence of vertex corrections to the conductivity is closely related to the momentum independence of the self-energy\cite{DMFT}.
Further, the presence of Umklapp processes, which allow electron-electron scattering to contribute to the resistivity in the pure limit\cite{Fukuyama}, is implicit in the above formula.

In a strongly correlated metal, $s$ may be approximated by its value in the unitary scattering limit\cite{MMV,Hewson}, $s_u=2n/3\pi D_0$, where $n$ is the conduction electron density and $D_0$ is the {\it bare} density of states (DOS) at the Fermi energy. In the low temperature, pure 
limit one finds (see the Methods section) that
\begin{eqnarray}
A = \frac{16 n k_B^2}{\pi \hbar e^2  \langle v_{0x}^2 \rangle D_0^2 \omega^{*2}}, \label{Akubo}
\end{eqnarray}
where $\langle \dots \rangle$ denotes an average over the Fermi surface.
Note that neither the DOS nor the Fermi velocity are renormalised in this expression. Indeed, all of the many-body effects are encapsulated by $\omega^*$, which determines the magnitude of the frequency dependent term in $\Sigma''(\omega,T)$, c.f. Eq. (\ref{eq:selfenergy}).

The Kramers-Kronig relation for the retarded self-energy\cite{luttinger61,KK} can be used to show (see Methods) that, in the pure limit, 
\begin{eqnarray}
\gamma = \gamma_0 \left( 1 - \frac{\partial \Sigma'}{\partial \omega}\right) = \gamma_0 \left( 1 + \frac{4 s_u\xi}{\pi  \omega^*} \right), 
\label{eqn:gamma}
\end{eqnarray}
where $\gamma_0 = {\pi^2}k_B^2 D_0/3$ is the linear coefficient of the specific heat for a gas of non-interacting fermions, $\Sigma'$ is the real part of the self energy and $\xi\approx1$ is a pure number defined in the Methods section.
Thus we see that the renormalisation of $\gamma$ is also controlled by $\omega^*$.  For a strongly correlated metal the effective mass, $m^*\gg m_0$, the bare (band) mass of the electron, hence $s_u \gg \omega^*$ and
$
\gamma 
\simeq (8 n k_B^2\xi 
)/(9 \omega^*)
$. The corrections to this approximation are given in the Methods section.

Combining the  above results we see that the KWR is
\begin{eqnarray}
 \frac{A}{\gamma^2}
=\frac{81}{4\pi \hbar k_B^2 e^2} \cdot \frac{ 1 }{\xi^2 n D_0^2 \langle v_{0x}^2 \rangle }.\label{eqn:KWRgen}
\end{eqnarray}
First, we note that in this ratio the dependence of the individual factors on $\omega^*$ has vanished. Hence the KWR is \emph{not renormalised}. On the other hand, while the first factor contains only fundamental constants, the second factor is clearly \emph{material dependent} as it depends on the electron density, the DOS and the Fermi velocity of the non-interacting system.  An important corollary to this result is that band-structure calculations should give accurate predictions of the KWR via Eq. (\ref{eqn:KWRgen}) as none of the properties on the right hand side are renormalised.

 The wide range of KWRs found in layered materials suggests that $\xi^2n D_0^2 \langle v_{0x}^2 \rangle$ varies significantly in these materials. For example, for highly anisotropic materials $\langle v_{0x}^2 \rangle$ may vary by more than an order of magnitude depending on which direction the resistivity is measured in. This effect needs to be taken into account  if we wish to understand what the KWR has to tell us about strongly correlated metals.  Further, Eq. (\ref{eqn:KWRgen}) suggests that the reason that the transition metals and the heavy fermions have `constant' KWRs is that $\xi^2n D_0^2 \langle v_{0x}^2 \rangle$ is roughly constant across each class of materials and that  $a_{HF}\ne a_{TM}$ because $\xi^2n D_0^2 \langle v_{0x}^2 \rangle$ is different in the  two different classes of materials.   Therefore, we propose that a more fundamental ratio is 
\begin{eqnarray}
\frac{A f_{dx}(n)}{\gamma^2}=\frac{81}{4\pi \hbar k_B^2 e^2},
\label{eqn:newRatioGen}
\end{eqnarray}
where $f_{dx}(n)\equiv n D_0^2 \langle v_{0x}^2 \rangle\xi^2$ may be written in terms of the dimensionality, $d$, of the system, the electron density and, in layered systems, the interlayer spacing or the interlayer hopping integral.

For simplicity we assume that  the reasonably isotropic materials (the heavy fermions, the transition metals and Rb$_3$C$_{60}$) are isotropic Fermi liquids and the layered materials (i.e., the transition metal oxides and organic charge transfer salts based on BEDT-TTF) have warped cylindrical Fermi surfaces; $f_{dx}(n)$ is derived for these band structures in the Methods section. This allows us to  test explicitly,  in Fig.~\ref{fig:KWR-new}, the  prediction of Eq. (\ref{eqn:newRatioGen})  against previously published experimental data 
 for a variety of strongly correlated metals.     It is clear that the new ratio is in good agreement with the data for {\it all} of the materials  investigated. 
We therefore see that the observations of constant KWRs for the heavy fermions and for the transition metals are due not only to the profound but also to the prosaic. The renormalisation of $\gamma^2$ cancels with that of $A$ due to the Kramers-Kronig relation for the self-energy; but the unrenormalised properties are remarkably consistent within each class of materials. Further, the large KWRs in transition metal oxides, the organics and UBe$_{13}$  are simply a consequence of the small values of $f_{dx}(n)$ in these materials.  
 Therefore, the absolute value of the KWR does not reveal anything about electronic correlations unless the material specific effects, described by $f_{dx}(n)$, are first accounted for. 
 
 It will be interesting to identify and understand strongly correlated metals that are not described by Eq. (\ref{eqn:newRatioGen}). Our calculation already gives some clues as to when this might happen: for example, when the self-energy is strongly momentum dependent or when the are significant vertex corrections to the conductivity. Another outstanding challenge is to understand the KWR in compensated semimetals\cite{Terashima,Chopra,Collan}.

\section*{METHODS}

\subsection*{Resistivity}

To calculate $A$ we  approximate the spectral density by
\begin{equation}
\mathcal{A}^2(\bm{k},\omega) \approx \frac{2 \pi Z
\delta(\omega-Z\xi_0(\bm{k}))}{-\Sigma''(\omega,T)} \label{A-approx}
\end{equation}
where $\xi_0(\bm{k})=\varepsilon_0(\bm{k})-\mu_0$ where $\mu_0$ is the
chemical potential of the bare system (i.e., in the absence of both electron-electron interactions and
impurity scattering) and $Z$ is the
quasi-particle weight defined as $Z^{-1}=1-(\partial/\partial
\omega)\Sigma'(\omega,0)|_{\omega=0}$. Eq. (\ref{A-approx}) will give the
right behavior in the limit $\Sigma''\to 0$ (Ref. \onlinecite{mahan}). Inserting Eq. (\ref{A-approx}) into
Eq. (\ref{eq:sigmakubo})  
and noting that at low temperatures $-\partial
f(\omega)/\partial \omega$ will be sharply peaked at $\omega=0$, we
replace $\omega$ by 0 in $\delta(\omega-Z\xi_0(\bm{k}))$. The
delta function can then be taken outside the $\omega$ integration, giving
\begin{equation}
\sigma_{xx}(T) \approx Z \hbar e^2  \langle v_{x0}^2
\rangle \int\frac{d\bm{k}}{(2\pi)^3} \delta(Z\xi_0(\bm{k}))
\int d\omega \frac{1}{-\Sigma''(\omega,T)} \left( \frac{-\partial
f(\omega)}{\partial \omega}\right).
\end{equation}
The $\bm{k}$-space integral here is (half)
the renormalised DOS, $D^*$, at the renormalised Fermi energy. As $D^*=D_0/Z$, where $D_0$ is the
DOS  of the bare system at its Fermi energy, we get
\begin{equation}
\sigma_{xx}(T)  = \hbar e^2  \langle v_{0x}^{2} \rangle D_0
\int d\omega \frac{\left( {-\partial f(\omega)}/{\partial \omega}\right)}{-2\Sigma''(\omega,T)}. \label{eq:mmveq3}
\end{equation}

Using ${-\partial f(\omega)}/{\partial \omega}\rightarrow\delta(\omega)$ as $T\rightarrow0$ it follows from
Eq. (\ref{eq:selfenergy}) 
that the zero temperature resistivity is given by
$\rho_0 = ( e^2  \tau_0 \langle v_{0x}^2 \rangle  D_0)^{-1}$. With the temperature dependence of the resistivity
given by $\rho(T)=\rho_0+AT^2$, \eq{mmveq3} yields
\begin{eqnarray}
AT^2 &=& \rho - \rho_0
 = \frac{1}{\hbar e^2  \langle v_{0x}^2 \rangle  D_0}
 \left( \left[ \int  d\omega  \frac{\left(-{\partial f(\omega)}/{\partial \omega}\right)}
 {-2 \Sigma''(\omega,T)} \right]^{-1} - \frac{\hbar}{\tau_0} \right). \label{eq:atsquared}
\end{eqnarray}

We now consider the pure limit, $\tau_0 \rightarrow \infty$. At sufficiently low temperatures
the contribution to the integral from the region $\omega>\omega^*$ is small, so it is a
good approximation to use Eq. (\ref{eq:selfenergy}) 
for $\Sigma''(\omega,T)$ for all $\omega$.
The resulting integral can then be evaluated analytically,
\begin{eqnarray}
\int^{\infty}_{-\infty}  d\omega\,  \frac{(-\partial f(\omega)/\partial \omega)}
{- 2\Sigma''(\omega,T)} \approx \frac{\omega^{*2}}{2s}\int_{-\infty}^{\infty}d\omega\,
\frac{(-\partial f(\omega)/\partial \omega)}{\omega^2+(\pi k_B T)^2} = \frac{(\omega^*/k_B T)^2}{24 s}.
\end{eqnarray}
Eq. (\ref{Akubo})  follows upon taking $s=s_u$.

\subsection*{Real part of the self energy}

In order to calculate $\gamma$ we need to know the real part of the self-energy. 
To evaluate this we apply the Kramers-Kronig relation\cite{luttinger61,KK} for the retarded self-energy, 
\begin{eqnarray}
\Sigma'(\omega,T) = \Sigma'(\infty,T) + \frac{1}{\pi} \mathcal{P} \int^{\infty}_{-\infty} \frac{\Sigma''(\omega',T)}{\omega'-\omega} d\omega' ,
\end{eqnarray}
where $\mathcal{P}$ indicates the principal part of the integral. 

For $T=0$ and taking the pure limit one finds that for $|\omega|\ll\omega^*$,
\begin{equation}
\frac{1}{\pi}{\mathcal P}\int_{-\infty}^{\infty}\frac{\Sigma''(\omega')}{\omega'-\omega}d\omega'=
-\frac{2s}{\pi}\left[y+\frac{1}{2}y^2 \ln\left|\frac{1-y}{1+y}\right|+\int_1^{\infty}dy'\,\frac{F(y')}{y'}
\sum_{n=0}^{\infty}\left(\frac{y}{y'}\right)^{2n+1}\right],
\end{equation}
where $y=\omega/\omega^*$, $y' = \omega'/\omega^*$, and we have used the fact that $F(y)$ is an even function. Therefore, in the limit $|y|\ll 1$
\begin{equation}
\Sigma'(\omega,0)= \Sigma'(\infty,0)-\frac{4 s_u \xi}{\pi}\frac{\omega}{\omega^*}  \label{eq:realse} +  \mathcal{O}\left(\frac{\omega^3}{\omega^{*3}}\right),
\end{equation}
where $1<2\xi \equiv 1+ \int_1^\infty y^{-2} F(y)  dy\leq 1+ \int_1^\infty y^{-2}  dy = 2$ as $F(y)\leq1$ for $y\geq 1$.   
Provided $F(y)$ decreases sufficiently slowly as $y\rightarrow\infty$, we expect $\xi\approx1$. In general, changes to the exact form of the self energy will simply lead to small changes in $\xi$ provided the boundary conditions for $\Sigma$ remain the same. 
Note that $\Sigma'(\infty,0)$ is just the shift in the  zero temperature chemical potential due to many-body interactions.

The form of the self energy given in Eq. (\ref{eq:selfenergy}) has a kink at $\omega=\omega^*$. Kinks are found in the self-energies of local Fermi liquid theories, such as dynamical mean-field theory\cite{Byczuk}. However, the location, and even the existence, of this kink is not important for our results. In order to calculate $A$ and $\gamma$ one must integrate over all $\omega$ (see above), therefore any sharp features in $\Sigma''(\omega,T)$ will be washed out. 

\subsection*{Band structure}

$A$, $\gamma$ and $n$ are relatively straightforward to determine experimentally. It is harder to directly measure $D_0$ and $\langle v_{0x}^2\rangle$. Therefore, we consider two model band structures. 
(i) For an isotropic Fermi liquid $\varepsilon_0(\bm{k})=\hbar^2\bm{k}^2/2m_0$, $D_0 ={m_0 k_F}/{\hbar^2 \pi^2}$ and
$\langle v_{0x}^2\rangle ={\hbar^2 k_F^2}/{3m_0^2}$, where  $k_F=\sqrt[3]{3\pi^2 n}$.
Hence, for  $\xi=1$,     $f_{3x}(n)=\sqrt[3]{3n^{7} /\pi^{4}\hbar^{6}} $. 
(ii) To study layered materials we employ the simple model dispersion $\varepsilon_0(\bm{k})=\hbar^2\bm{k}_{ab}^2/2m_0-2t_{\perp0}\cos ck_\perp$, where $\bm{k}=(\bm{k}_{ab},k_\perp)$, $\bm{k}_{ab}$ is the in-plane wavevector, $k_\perp$ is the wavenumber perpendicular to the plane, $c$ is the interlayer spacing and  $t_{\perp0}$ is the bare interlayer hopping integral. 
If $A$ is taken from measurements of the resistivity parallel to the plane $\langle v_{0\|}^2\rangle ={\hbar^2k_{F}^2}/2m_0^2$. 
But, for $A$ measured perpendicular to the plane 
$\langle v_{0\perp}^2\rangle ={2c^2 t_{\perp0}^2}/{\hbar^2}$. In either case for the warped cylindrical Fermi surface we are now considering $D_0 = m_0/\pi c \hbar^2$ and $k_F=\sqrt{2 \pi c n}$. So, for  $\xi=1$, we find that $f_{2\|}(n) = n^2 /\pi c \hbar^2$ and $f_{2\perp}(n) = 2n m_0^2 t_{\perp0}^2 / \pi^2  \hbar^6$.

This formalism can straightforwardly be generalised to include other factors known to affect the KWR by extending the definition of $f_{dx}(n)$. For example, Kontani\cite{Kontani} has shown that in the $N$-fold orbitally degenerate periodic Anderson model $A/\gamma^2\propto[N(N-1)]^{-1}$, in good agreement with experiments on heavy fermions with orbital degeneracy\cite{Tsujii,Canfield}. This result is specific to this particular model and the systems so far studied\cite{Tsujii,Canfield} all have rather similar values of $\xi^2n D_0^2 \langle v_{0x}^2 \rangle$. However,  it is clear, from a comparison of Kontani's\cite{Kontani} calculation with ours, that if orbitally degenerate systems with different electron densities or reduced dimensionalites were fabricated $f_{dx}(n)$ will need to be included  to understand the relationship between $A$ and $\gamma$. Hussey\cite{Hussey} has shown that the number of sheets of the Fermi surface also affects the KWR. It is straightforward to generalise the above calculations of $f_{dx}(n)$ to  Fermi surfaces with any number of sheets. Finally, we note that if one relaxes the condition $m^*\gg m_0$ then one finds that 
${A f_{dx}(n)}/{\gamma^2}=(81/4\pi \hbar k_B^2 e^2)(1-m_0/m^*)^2$, i.e., the ratio vanishes as $m^*\rightarrow m_0$. It is therefore perhaps somewhat surprising that a constant KWR is seen in the transition metals, which do not all have such large effective masses as the other materials discussed above.

{\bf Acknowledgements.}
It is a pleasure to thank Javier Castro, Nigel Hussey, Malcolm Kennett, Ross McKenzie, Jaime Merino, Mike Smith, Tom Stace, Chandra Varma, Andrew White and Jochem Wosnitza for their helpful comments. This research was supported under the Australian Research Council's (ARC) Discovery Projects funding scheme (project DP0878523). B.J.P.  is the recipient of  an ARC Queen
Elizabeth II Fellowship (DP0878523).

{\bf Author contributions.}
This project was planned and led by B.J.P. All authors contributed equally to the derivation. The analysis of the previously published experimental data was carried out by A.C.J. The paper was written by B.J.P. with significant input from A.C.J. and J.O.F.

\end{document}